# Enhancement of the superconducting transition temperature of FeSe by intercalation of a molecular spacer layer.

Matthew Burrard-Lucas,[a] David G. Free,[a] Stefan J Sedlmaier,[a] Jack D Wright,[b] Simon J Cassidy,[a] Yoshiaki Hara,[a] Alex J Corkett,[a] Tom Lancaster,[c] Peter J Baker,[d] Stephen J Blundell[b] and Simon J Clarke.[a]

[a]Department of Chemistry, University of Oxford, Inorganic Chemistry Laboratory, South Parks Road, Oxford OX1 3QR, UK.

[b]Department of Physics, University of Oxford, Clarendon Laboratory, Parks Road, Oxford OX1 3PU, UK.

[c]Department of Physics, Durham University, South Road, Durham DH1 3LE, UK

[d]ISIS facility, Rutherford Appleton Laboratory, Chilton, Oxon, OX11 0QX, UK

**The recent discovery of high temperature superconductivity in a layered iron arsenide[1] has led to an intensive search to optimize the superconducting properties of iron-based superconductors by changing the chemical composition of the spacer layer that is inserted between adjacent anionic iron arsenide layers[2-7]. Until now, superconductivity has only been found in compounds with a cationic spacer layer consisting of metal ions: $Li^+$, $Na^+$, $K^+$, $Ba^{2+}$ or a PbO-type or perovskite-type oxide layer. Electronic doping is usually necessary to control the fine balance between antiferromagnetism and superconductivity. Superconductivity has also been reported[8] in FeSe, which contains neutral layers similar in structure to those found in the iron arsenides but without the spacer layer. Here we demonstrate the synthesis of $Li_x(NH_2)_y(NH_3)_{1-y}Fe_2Se_2$ ($x$ ~0.6 ; $y$ ~ 0.2), with lithium ions, lithium amide and ammonia acting as the spacer layer, which exhibits superconductivity at 43(1) K, higher than in any FeSe-derived compound reported so far and four times higher at ambient pressure than the transition temperature, $T_c$, of the parent $Fe_{1.01}Se$. We have determined the crystal structure using neutron powder diffraction and used magnetometry and muon-spin rotation data to determine the superconducting properties. This new synthetic route opens up the possibility of further exploitation of related molecular intercalations in this and other systems in order to greatly optimize the superconducting properties in this family.**

The tetragonal phase of $Fe_{1+\delta}Se$ (0.01 ≤ $\delta$ ≤ 0.04) with the anti-PbO-type crystal structure displays superconductivity ($T_c$ ~ 8.5 K) when $\delta$ = 0.01 and the superconductivity is destroyed by additional interstitial Fe.[9] The compound bears close structural resemblance to LiFeAs[10] (both compounds contain $FeQ_4$ (Q = Se, As) tetrahedra that are highly compressed in the basal plane compared with other iron-based superconductors) and both compounds differ from the canonical iron-based superconducting system in that they superconduct when as close to stoichiometric as possible (i.e. when they formally contain $Fe^{2+}$) and do not require chemical substitution to drive them away from the itinerant antiferromagnetic state into the superconducting regime (compare, for example $LnFeAsO$,[1] $BaFe_2As_2$[6] and $NaFeAs$[11]). Under an applied hydrostatic pressure the $T_c$ of $Fe_{1.01}Se$ increases to 37 K at 7 GPa.[12,13] $Fe_{1+\delta}Se$ seems less inviting than the arsenide systems in terms of chemical flexibility. For example as part of this work we attempted to insert Li using excess *n*-BuLi in hexane at room temperature but this results in the reduction of iron to the elemental state with formation of $Li_2Se$ and Fe as the only crystalline products. Relatives of $Fe_{1+\delta}Se$ in which alkali metals separate the layers are the subject of much recent research and remain controversial. The lack of much redox flexibility in these iron selenide systems synthesised at high temperatures is underlined by the fact that the compositions are close to the "245" stoichiometry of $K_{0.8}Fe_{1.6}Se_2$ (with a highly defective variant of the $ThCr_2Si_2$ structure) where Fe is again formally divalent. These systems are characterised by a Fe/vacancy ordered[14] antiferromagnetic state which appears not to support superconductivity and differs from the antiferromagnetic state of the iron arsenide parent systems in that the magnetic structure[15] is related to the iron/vacancy ordering scheme and the ordered moment localised on Fe is around 3.4 $\mu_B$[15] as opposed to < 1$\mu_B$ in the arsenides. The crystal structures of these phases close to $K_{0.8}Fe_{1.6}Se_2$ are complex and there is reported evidence for coexistence of regions with different Fe/vacancy ratios and different ordering schemes in the same sample.[16] Phases of this type with variously reported compositions show bulk superconductivity, although this is often difficult to reproduce[14] and it has recently been suggested from NMR studies that the superconducting samples in the Rb-Fe-Se phase field consist of intergrowths of antiferromagnetic and non-superconducting $Rb_{0.8}Fe_{1.6}Se_2$ and regions of superconducting material which are electron-doped and close in composition to $Rb_{0.3(1)}Fe_2Se_2$.[17] Whether such a superconducting phase can be prepared pure is the subject of ongoing research.

A question arises as to how to make compounds in which stoichiometric iron selenide layers, which evidently support superconductivity in tetragonal $Fe_{1+\delta}Se$ itself, may be separated by intervening layers as is possible in the case of the formally mono-anionic FeAs layers in other iron-based superconductors. Nature provides a clue in the compound tochilinite in which close-to-stoichiometric FeS layers (like those in the mackinawite polymorph of FeS) are separated by brucite-type $Mg(OH)_2$ layers to form $(FeS)(Mg(OH)_2)_y$ (y ~ 0.8 – 1).[18] In this letter we demonstrate that the reaction of $Fe_{1+\delta}Se$ with a solution of lithium in liquid ammonia (see experimental methods section) is a way to simultaneously intercalate both lithium cations, amide anions and ammonia molecules between the FeSe layers to produce $Li_x(NH_2)_y(NH_3)_{1-y}Fe_2Se_2$ (x ~0.6; y ~ 0.2). The result is a dramatic enhancement of $T_c$ which reaches 43(1) K. These properties are in line with a recent report[19] of a series of nominal composition $A_xFe_2Se_2$ (A = ammonia-soluble electropositive metal) including a lithium intercalate which we suspect also contains amide and ammonia. Here we report on the crystal structure and superconducting properties of products obtained by the intercalation reactions using from $Li/NH_3$ and $Li/ND_3$ solutions. Refinement of a model for the deuterated compound against neutron powder diffraction data shows that amide ions and ammonia molecules occupy the

8-coordinate sites in the ThCr$_2$Si$_2$ structure type and are coordinated by Li ions. There is evidence for weak N–H⋯Se hydrogen bonds as found for ammonia intercalates of TiS$_2$.[20]

The lithium/ammonia solutions were rapidly decolourised by Fe$_{1+\delta}$Se at −78 °C. This is consistent either with the classic method for decomposing the metastable solution of solvated electrons using a "rusty nail" as a catalyst for the formation of lithium amide and hydrogen, or it indicates. Donation of the solvated electrons from the alkali metal ammonia solution to empty bands in the solid with Li ions co-inserted to balance the charge in a reductive intercalation reaction. The product was a black powder with a much finer grain size than the parent Fe$_{1+\delta}$Se material. The products were extremely air sensitive. X-ray powder diffraction (XRPD) showed no evidence for the starting material or other products above the 5% level and the diffraction peaks were indexed on a body centred tetragonal unit cell with lattice parameters $a$ = 3.8249(2) Å and $c$ = 16.5266(9) Å at room temperature. The basal lattice parameter, $a$ (= √2 × Fe–Fe distance) is 1.4% larger than the value of 3.7734(1) Å reported for Fe$_{1.01}$Se.[9] The $c$ lattice parameter of 16.5266(9) Å is almost 20 % larger than the value found in K$_{0.8}$Fe$_{1.6}$Se$_2$[14] with the ThCr$_2$Si$_2$ structure type. This suggests that more than just a metal cation has been inserted between the layers. Indeed, a preliminary structural model with Li located on the 8-coordinate site surrounded by selenide ions had unreasonably long Li–Se distances. (This structure is known in, for example, LiCu$_2$P$_2$,[21] but this compound accommodates lithium in a relatively small 8-coordinate site resulting from the presence of short P–P bonds). Rietveld refinement[22] against laboratory XRPD data of the scattering from the 8-coordinate site suggested a chemically unreasonable formula NFe$_2$Se$_2$ suggesting that light Li and/or H atoms were also present. A control reaction in which Fe$_{1+\delta}$Se was stirred in ammonia solution with no lithium present resulted in no change in the XRPD pattern.

Neutron powder diffraction (NPD) patterns were collected from samples synthesised using 0.5 moles of Li per mole of FeSe with either NH$_3$ or ND$_3$ as solvent. The XRPD patterns of these products were similar, but their NPD patterns were dramatically different (the intensities varied greatly between the two compounds because of the greatly different neutron scattering lengths for H (−3.74 fm) and D (+6.67 fm), and the H-containing material produced a characteristic incoherent background) proving that the samples contain H(D). A structural model was obtained from the deuterated sample by starting from the model suggested by the X-ray refinements with N in the site 8-coordinate by Se and computing Fourier difference maps to reveal the remaining nuclear scattering density. Refinements against data from the GEM diffractometer at room temperature and the HRPD diffractometer at 8 K on the same sample of deuterated material produced similar structural models at the two temperatures. The initial assumption of a formula (LiND$_2$)Fe$_2$Se$_2$ resulted in an apparently satisfactory fit to the low temperature HRPD data (which emphasises the short $d$-spacing data) using a model in which the D atoms were located on crystallographic positions (16$m$ site: ($x, x, z$)) which refined freely to be about 1 Å from the N atom (2$a$ site: (0, 0, 0)) and with the N–D bonds directed towards the selenide anions. In this initial model the D site had a site occupancy fixed at one quarter resulting in one ND$_2$ moiety per square prismatic site. Li ions were located at sites (0, ½, 0) (4$c$ site) between the ND$_2$ moieties but with very large displacement ellipsoids suggestive of disorder in the intercalant. This model proved unsatisfactory when the ambient temperature GEM data were investigated. In particular the 002 reflection which accidently has zero intensity in the deuterated material could only be modelled accurately when further D atoms were included in the 8$i$ sites ($x$, 0, 0) in the same plane as the N atoms and which freely refined to be about 1 Å from the N atom.

Further Li was located at the 2*b* site (0, 0, ½). Although Li makes a minor contribution to the scattering in the presence of N, D, Fe and Se, the Li site occupancy was consistently less than the formula (LiND$_2$)Fe$_2$Se$_2$ would suggest. Unconstrained refinement of the site occupancies of the two D sites and the Li sites produced a refined composition of Li$_{0.6(1)}$ND$_{2.8(1)}$Fe$_2$Se$_2$ which may be reformulated Li$_{0.6(1)}$(ND$_2$)$_{0.2(1)}$(ND$_3$)$_{0.8(1)}$Fe$_2$Se$_2$ with intercalation of lithium amide and ammonia. About 20 % of the Li included in the synthesis appears as a separate LiND$_2$ phase present in the products. Re-examination of the HRPD data collected at 8 K resulted in a significant improvement in the fit when ND$_3$ was accommodated in place of some LiND$_2$ and the refined composition at 8 K using HRPD data was similar to that obtained from the refinement against GEM data at 298 K with some redistribution of Li. The final model is shown in Figure 1 and the supporting information. In the model N–D bonds of about 1 Å from [ND$_2$]$^-$ and ND$_3$ species are directed towards the selenide ions with D···Se distances of 2.75 Å consistent with weak hydrogen bonding interactions comparable to those found in the lithium/ammonia intercalates of TiS$_2$.[20] The uncertainty in our refinements is partly associated with the large displacement ellipsoids for the intercalated species typical for similar systems,[20,23] but the refinements show that most of the N is present in ND$_3$ molecules and the Li : amide ratio exceeds unity implying donation of electrons (0.2(1) per FeSe unit) to the FeSe layers, consistent with the proposed Rb$_{0.3(1)}$Fe$_2$Se$_2$ superconducting phase suggested by NMR measurements.[17] The crystal structure obtained from the refinement against NPD data is shown in Figure 1 and the Rietveld fits are shown in Figure 2.

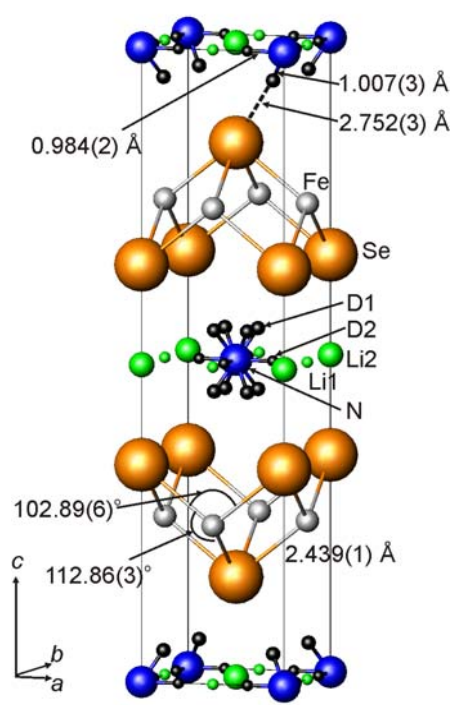

**Figure 1.** The crystal structure obtained from the refinement against neutron powder diffraction data on Li$_{0.6(1)}$(ND$_2$)$_{0.2(1)}$(ND$_3$)$_{0.8(1)}$Fe$_2$Se$_2$ at 298 K (GEM data). In the model each square prism of Se atoms contains either an [ND$_2$]$^-$ anion or an ND$_3$ molecule and these are both modelled as disordered over four orientations. The sizes of the spheres representing the Li atoms are in proportion to their site occupancies.

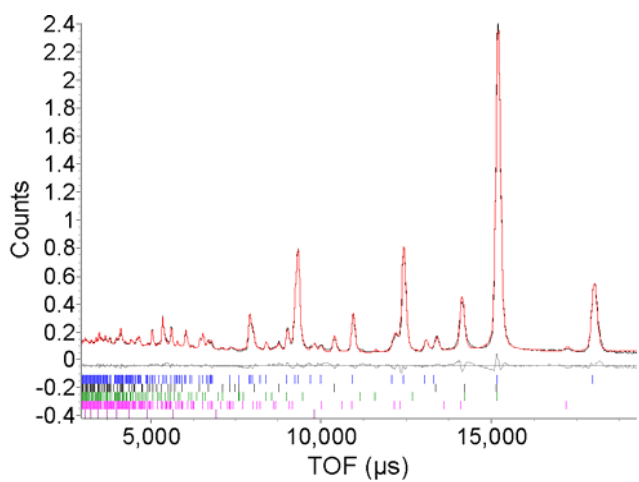

**Figure 2.** Rietveld refinement against GEM data for $Li_{0.6(1)}(ND_2)_{0.2(1)}(ND_3)_{0.8(1)}Fe_2Se_2$ at 298 K. The data are from the $2\theta = 64.6°$ detector bank. Allowed peak positions are marked by vertical lines: from top the main phase and minority phases: FeSe (hexagonal, present in the starting material) (10% by mass), FeSe (tetragonal) (1.2 %), $LiNH_2$ (1.8 %), and Fe (1.5 %).

Compared with tetragonal $Fe_{1+\delta}Se$ the structure is expanded in the basal direction by 1.4 %. At ambient temperatures the extremely compressed $FeSe_4$ tetrahedra found in $Fe_{1+\delta}Se$ are retained with Se–Fe–Se bond angles (ambient temperature) of 102.89(6)° (× 2) and 112.86(3)° (× 4) compared with values of 103.9° (× 2) and 112.3° (× 4) found for $Fe_{1+\delta}Se$.[9] The Fe–Se bond distances are 2.3958 Å in $Fe_{1+\delta}Se$ and these are significantly larger in the intercalates reaching 2.439(1) Å in $Li_{0.6(1)}(ND_2)_{0.2(1)}(ND_3)_{0.8(1)}Fe_2Se_2$ at ambient temperature, consistent with electron doping of the system. At 8 K the Se–Fe–Se bond angles of 104.40(8)° (× 2) and 112.06(4)° (× 4) indicate a very slight decrease in the basal-direction compression of the $FeSe_4$ tetrahedra and the Fe–Se bonds contract to 2.408(1) Å. The high resolution data obtained on HRPD show that unlike in $Fe_{1+\delta}Se$ [24] there is no structural distortion evident down to 8 K.

SQUID magnetometry measurements (Figure 3) made under zero-field-cooled (ZFC) and field-cooled (FC) conditions showed that the products of the reactions with lithium ammonia solutions exhibit bulk superconductivity (superconducting volume fractions of about 50 %) with $T_c$ = 43(1) K. There is no discernible isotope effect and $T_c$ is similar for all the intercalates of FeSe within the experimental uncertainty. There are some variations between samples in the volume fraction based on the size of the diamagnetic signal and in the size of the normal state susceptibility. Measurements of magnetisation as a function of field showed the presence of a ferromagnetic impurity corresponding to about 1-2 % by mass of elemental Fe which explains the large normal state susceptibility. The Fe may originate from the Fe ions present in the interlamellar space in the $Fe_{1+\delta}Se$ parent material[9] and reductively extruded by the reducing solution.

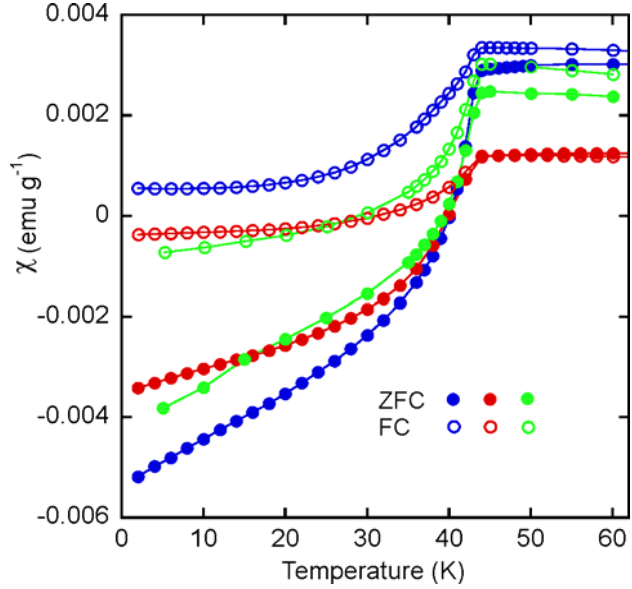

**Figure 3.** Magnetic susceptibility measurements (zero-field-cooled (ZFC) and field-cooled (FC)) on the three samples of Li$_{0.6(1)}$(NH/D$_2$)$_{0.2(1)}$(NH/D$_3$)$_{0.8(1)}$Fe$_2$Se$_2$ measured using μSR and neutron powder diffraction (NPD). Red symbols: μSR sample; blue symbols: H-containing NPD sample; green symbols: D-containing NPD sample.

We have used muon-spin rotation (μSR) to measure the increase in $B_{rms}$, the rms width of the magnetic field distribution, due to the development of the superconducting vortex lattice below $T_c$. These results suggest a superconducting volume fraction in the range 40-50 % for the sample which shows the smallest volume fraction in the magnetic susceptibility measurements (Figure 3). Figure 4 (a) shows $B_{rms}$ as a function of temperature measured using a transverse field of 10 mT and the fitted curve is the result of three contributions (dashed lines) summed in quadrature, a temperature-independent normal-state contribution and two temperature-dependent contributions which account for the effect of superconductivity (and is similar to the two-gap behavior observed in Fe$_{1+\delta}$Se in Ref 25.) Using the proportionality between $B_{rms}$ and $1/\lambda_{ab}^2$, where $\lambda_{ab}=3^{1/4}\lambda$ is the in-plane penetration depth we can extract an estimate of $\lambda_{ab}$. This value places Li$_{0.6(1)}$(NH$_2$)$_{0.2(1)}$(NH$_3$)$_{0.8(1)}$Fe$_2$Se$_2$ on the main scaling line in a Uemura plot[26] of $T_c$ against superfluid stiffness $\rho_s=c^2/\lambda_{ab}^2$ (Figure 4(b)), in common with underdoped cuprates and many other iron-based superconductors.

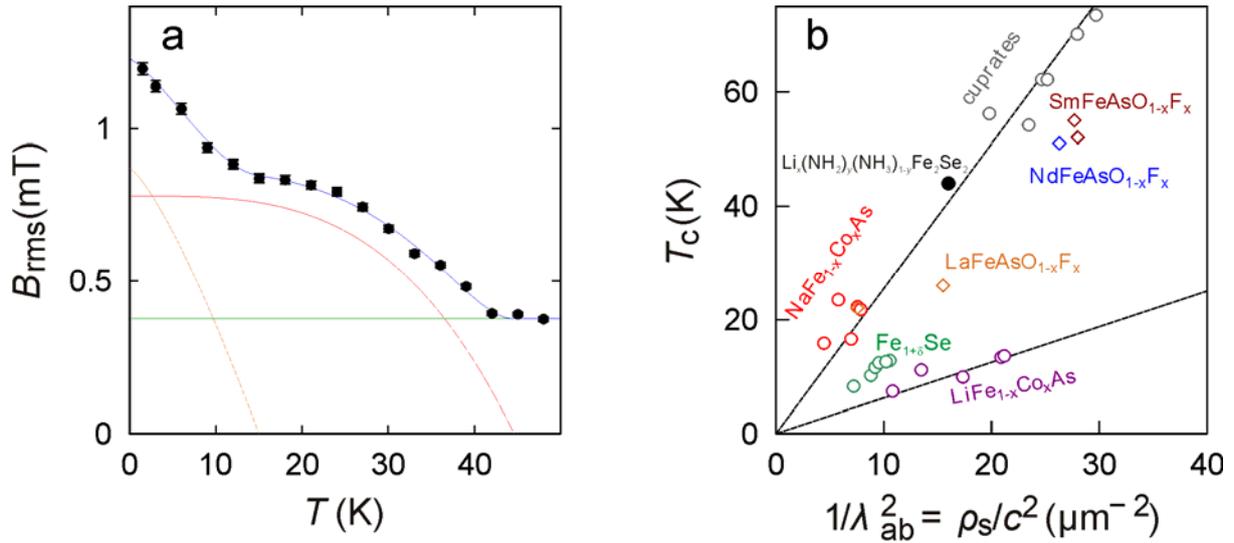

**Figure 4**: (a) Field width $B_{rms}$ as a function of temperature. The fitted line comprises the three dashed contributions summed in quadrature. (b) A Uemura plot of $T_c$ against superfluid stiffness $\rho_s = c^2/\lambda_{ab}^2$ showing that $Li_{0.6(1)}(NH_2)_{0.2(1)}(NH_3)_{0.8(1)}Fe_2Se_2$ falls on the main scaling line. Data for $Fe_{1+\delta}Se$ are shown for different pressures (Ref. 25) and the plot is adapted from Ref. 10.

These superconductors $Li_{0.6(1)}(NH_2)_{0.2(1)}(NH_3)_{0.8(1)}Fe_2Se_2$ are thermally unstable: gentle heating of the samples under vacuum at below 100 °C was sufficient to decompose the intercalates. The powder diffraction pattern of the decomposition products of $Li_{0.6(1)}(NH_2)_{0.2(1)}(NH_3)_{0.8(1)}Fe_2Se_2$ at 400 °C revealed FeSe and $Li_2Se$ as the dominant crystalline products (see supplementary information). Thermogravimetric analysis revealed two mass loss features below 250 °C (see supplementary information). These are consistent with the facile loss of 0.75(2) moles of $NH_3$ per mole of $Li_{0.6(1)}(NH_2)_{0.2(1)}(NH_3)_{0.8(1)}Fe_2Se_2$ below 100 °C suggesting that the intercalated ammonia is lost readily and subsequent decomposition occurs above this temperatures of the intercalated lithium amide to form gaseous species and $Li_2Se$. The facile loss of intercalated ammonia occurs in a similar temperature range in the intercalates of $TiS_2$.[20]

In summary we have identified the products of the reaction between tetragonal $Fe_{1+\delta}Se$ and lithium/ammonia solutions as intercalates in which lithium ions, amide ions and the molecule $NH_3$ occupy sites between the FeSe layers which become well separated compared with $Fe_{1+\delta}Se$ itself and in which superconductivity occurs with a $T_c$ of 43(1) K. The refined composition and the increase in Fe–Se bond lengths suggest electronic doping of the FeSe layers. The dramatic change in the electronic properties may result partially from the increase in the separation of the layers. The reducing reaction conditions would be expected to completely remove the few % of interstitial Fe ions that are found to have a very detrimental effect on the superconducting properties of $Fe_{1.01}Se$,[9] and this change may also contribute to the observation of the enhanced superconducting properties. Measurement of the superconducting state using μSR spectroscopy shows features similar to those in $Fe_{1+\delta}Se$ and the superfluid stiffness is found to follow the scaling behaviour of Uemura that is valid for many classes of unconventional superconductor. Our results strongly imply that incorporation of salts and molecular groups into spacer layers in iron arsenides and selenides in intercalation reactions is a powerful new synthetic strategy that holds great promise for discovering new superconducting compounds.

**Experimental Methods.**

**Synthesis.** All manipulations were carried out in a Glovebox Technology Ltd argon-filled dry box with a $O_2$ and $H_2O$ content below 1 ppm or on a Schlenk line. Tetragonal FeSe was synthesised by grinding together high purity iron powder (ALFA 99.995 %) and selenium shot (ALFA 99.99 %) and heating them in a sealed silica tube to 700 °C at 2 °Cmin$^{-1}$, holding the temperature for 24 hours, cooling at 4°C min$^{-1}$ to 400 °C and holding this temperature for 24 hours before quenching the sample in water.

To synthesise the intercalates, FeSe and an amount of Li metal (Alrich 99 %) corresponding to the stoichiometry Li$_{0.5}$FeSe was placed in a Schlenk tube with a magnetic stirrer. This vessel and a cylinder of ammonia (BOC 99.98 %) were attached to a Schlenk line equipped with a mercury manometer. The Schlenk tube and the pipework extending to the regulator (previously flushed with ammonia) attached to the ammonia cylinder were evacuated. The Schlenk tube was placed in a bath of dry ice/ isopropanol (−78 °C) and when cooled, the Schlenk line was isolated from the vacuum pump. The valves on the ammonia cylinder and regulator were then opened allowing ammonia to condense onto the reactants. The ammonia pressure in the line was monitored using the mercury manometer and did not exceed 1 atm during the condensation process. When working on the 2 g scale approximately 50 cm$^3$ of NH$_3$ were condensed and the ammonia cylinder and regulator were then closed. The solution was observed to turn blue characteristic of solvated electrons but after about 30 minutes of stirring at −78 °C the blue colour was not evident. The Schlenk tube was allowed to warm with the slush bath and the ammonia and any evolved gases were allowed to evaporate out through the mercury manometer. A similar procedure was adopted when ND$_3$ was used except that the more valuable ND$_3$ was recondensed into its storage vessel as it evaporated from the reaction vessel.

**Structural characterisation.** Laboratory X-ray powder diffraction measurements (XRPD) were made using a Philips PW1730 (Cu$K\alpha_1$/$K\alpha_2$) or a Panalytical X'pert PRO instrument (Cu$K\alpha_1$) radiation. Neutron powder diffraction measurements on both hydrogen and deuterium containing compounds were made using the time of flight diffractometer HRPD (ambient temperature and 8 K) and the GEM diffractometer (ambient temperature) at the ISIS facility, Rutherford Appleton Laboratory, UK. On HRPD the samples were cooled in a closed cycle refrigerator. The samples were contained in 6mm diameter thin-walled vanadium cans which were sealed with indium gaskets.

**Magnetometry.** Magnetic susceptibility measurements were conducted using a Quantum Design MPMS-XL SQUID magnetometer. The powder samples of about 10 mg in mass were immobilized in gelatin capsules. Measurements were made in DC fields of 20 – 50 Oe in the temperature range 2 – 300 K after cooling in zero applied field (zero-field-cooled: ZFC) and in the measuring field (field-cooled: FC).

**Muon-spin rotation spectroscopy.** The μSR experiments were carried out at the ISIS Pulsed Muon Facility, UK. In a μSR experiment, spin-polarized muons are implanted in the bulk of a material and the time-dependence of their polarization monitored by recording the angular distribution of the subsequent positron decay. A sample of 0.5-1.0 g was packed into a square Ag-foil packet (2 cm square, ~25 μm foil thickness), then crimped tightly closed, and then mounted on a silver backing plate inside a He$^4$ cryostat. For the transverse field measurements, the sample was mounted on a haematite backing plate so that any muons missing the sample did not contribute to the precessing amplitude. In these experiments, muons stopped at random positions in the vortex lattice where they precessed about the total local magnetic field $B$ at the muon sites with frequency $\gamma_\mu B/2\pi$, where $\gamma_\mu/2\pi = 135.5$ MHz T$^{-1}$. The observed property of the experiment is the time evolution of the muon spin polarization $P_x(t)$, which is related to the distribution $p(B)$ of fields in the sample by

$$P_x(t) = \int_0^\infty p(B)\cos(\gamma_\mu Bt + \phi)dB$$

where the phase $\phi$ results from the detector geometry.

**Acknowledgements.** We are grateful to the ISIS facility including the GEM Xpress service for access to neutron and muon instruments and we acknowledge financial support from the UK EPSRC (Grant EP/I017844/1) and STFC (Grant EP/G067481/1).

**Contributions of co-authors.** MB-L, DGF, YH and SJS prepared the samples, DGF, AJC, SJS and SJC (Clarke) performed the diffraction data collection and structural analysis. JDW, TL, PJB and SJB performed the µSR measurements, MB-L, SJC (Cassidy), AJC and SJC (Clarke) performed the magnetometry and other characterisation measurements. SJC (Clarke) conceived the project and, with SJB, wrote the manuscript.

**References.**

# Enhancement of the superconducting transition temperature of FeSe by intercalation of a molecular spacer layer.

Matthew Burrard-Lucas,[a] David G. Free,[a] Stefan J Sedlmaier,[a] Jack D Wright,[b] Simon J Cassidy,[a] Yoshiaki Hara,[a] Alex J Corkett,[a] Tom Lancaster,[c] Peter J Baker,[d] Stephen J Blundell[b] and Simon J Clarke.[a]

[a]Department of Chemistry, University of Oxford, Inorganic Chemistry Laboratory, South Parks Road, Oxford OX1 3QR, UK.

[b]Department of Physics, University of Oxford, Clarendon Laboratory, Parks Road, Oxford OX1 3PU, UK.

[c]Department of Physics, Durham University, South Road, Durham DH1 3LE, UK

[d]ISIS Facility, Rutherford Appleton Laboratory, Chilton, Oxon, OX11 0QX, UK

**Supplementary Information**

**Figure S1.** Rietveld refinements against HRPD data and GEM data for Li$_{0.6(1)}$(ND$_2$)$_{0.2(1)}$(ND$_3$)$_{0.8(1)}$Fe$_2$Se$_2$ at 8K and 298K respectively. (a) HRPD data (blue) are (top to bottom) from the 168°, 90° and 35° data banks. The D-containing materials are shown on the right three panels and the data are compared with the H-containing materials (left three panels) showing the intensity differences due to the H/D contrast. (b) GEM data (black) run from, left to right and top to bottom, the 9°, 18°, 35°, 64°, 91° and 154° data banks respectively.

(a)

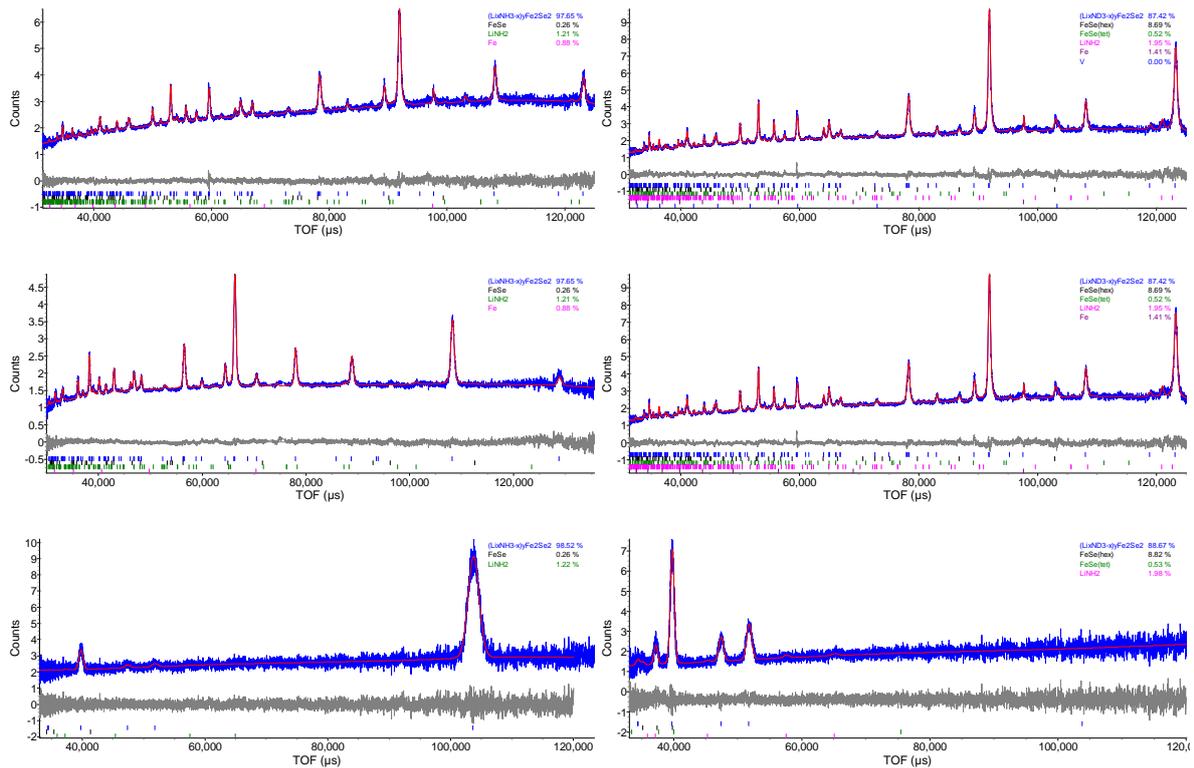

(b)
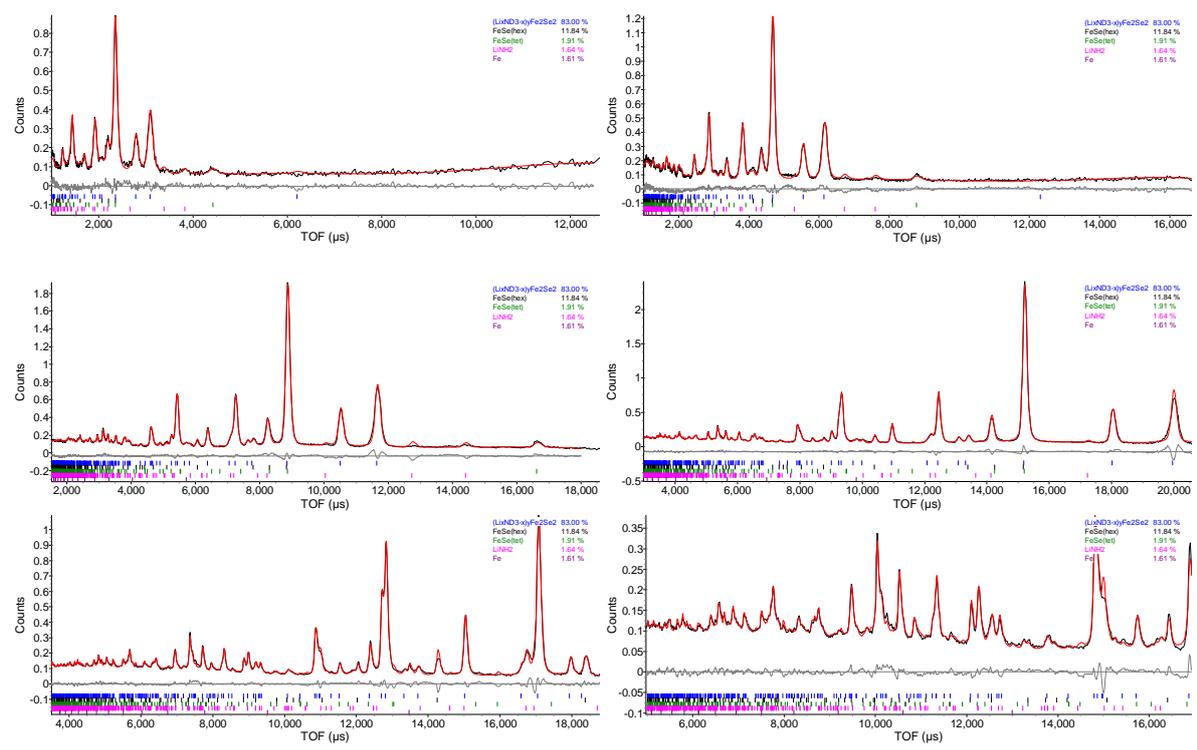

**Table S1.** Refined atomic coordinates for $Li_{0.6(1)}(ND_2)_{0.2(1)}(ND_3)_{0.8(1)}Fe_2Se_2$

298K GEM data

| Atom | Wyckoff Symbol | x | y | z | $U_{ij}$ (100×Å²) | | | Occ. |
| --- | --- | --- | --- | --- | --- | --- | --- | --- |
| | | | | | $u_{11}$ | $u_{22}$ | $u_{33}$ | |
| Fe | 4d | 0 | ½ | ¼ | 0.33(5) | $=u_{11}$ | 4.5(2) | 1 |
| Se | 4e | 0 | 0 | 0.3423(1) | 0.52(6) | $=u_{11}$ | 1.9(2) | 1 |
| N | 2a | 0 | 0 | 0 | 6.6(1) | $=u_{11}$ | $=u_{11}$ | 1 |
| D1 | 16m | 0.1158(7) | =x | 0.0480(1) | 6.6(1) | $=u_{11}$ | $=u_{11}$ | 0.25(2) |
| D2 | 8i | 0.2579(5) | 0 | 0 | 6.6(1) | $=u_{11}$ | $=u_{11}$ | 0.19(1) |
| Li1 | 4c | 0 | ½ | 0 | 6.6(1) | $=u_{11}$ | $=u_{11}$ | 0.15(2) |
| Li2 | 2b | 0 | 0 | ½ | 6.6(1) | $=u_{11}$ | $=u_{11}$ | 0.33(1) |

8K HRPD data

| Atom | Wyckoff Symbol | x | y | z | $U_{ij}$ (100×Å²) | | | Occ. |
| --- | --- | --- | --- | --- | --- | --- | --- | --- |
| | | | | | $u_{11}$ | $u_{22}$ | $u_{33}$ | |
| Fe | 4d | 0 | ½ | ¼ | 0.00(6) | $=u_{11}$ | 3.3(2) | 1 |
| Se | 4e | 0 | 0 | 0.3412(1) | 0.00(7) | $=u_{11}$ | 1.4(2) | 1 |
| N | 2a | 0 | 0 | 0 | 5.9(1) | $=u_{11}$ | $=u_{11}$ | 1 |
| D1 | 16m | 0.1245(6) | =x | 0.0458(2) | 5.9(1) | $=u_{11}$ | $=u_{11}$ | 0.297(3) |
| D2 | 8i | 0.2610(7) | 0 | 0 | 5.9(1) | $=u_{11}$ | $=u_{11}$ | 0.164(6) |
| Li1 | 4c | 0 | ½ | 0 | 5.9(1) | $=u_{11}$ | $=u_{11}$ | 0.236(2) |
| Li2 | 2b | 0 | 0 | ½ | 5.9(1) | $=u_{11}$ | $=u_{11}$ | 0.08(2) |

The large thermal displacements for N, D and Li, typical in these systems were constrained to be isotropic and equal.

**Table S2.** Refined structural parameters for Li$_{0.6(1)}$(ND$_2$)$_{0.2(1)}$(ND$_3$)$_{0.8(1)}$Fe$_2$Se$_2$

| Temperature (K) | 298 | 8 |
|---|---|---|
| Instrument | GEM | HRPD |
| Space Group | *I*4/*mmm* (no. 139) | *I*4/*mmm* (no. 139) |
| Z | 2 | 2 |
| *a* (Å) | 3.8149(2) | 3.8059(1) |
| *c* (Å) | 16.480(1) | 16.1795(6) |
| *V* (Å$^3$) | 239.84(3) | 234.35(2) |
| *c*/*a* | 4.3200(3) | 4.2512(2) |
| N–H/D(1) (Å) | 1.007(3) | 0.998(3) |
| N–H/D(2) (Å) | 0.984(2) | 0.993(3) |
| H/D(1)–N–H/D(1) (°) [2] | 103.4(4) | 95.7(3) |
| H/D(1)–N–H/D(2) (°) [2] | 116.0(1) | 118.3(9) |
| Fe–Se (Å) | 2.439(1) | 2.408(1) |
| Se–Fe–Se (°) [2] | 102.89(6) | 104.40(8) |
| Se–Fe–Se (°) [4] | 112.86(3) | 112.06(4) |
| H/D(1)⋯Se (Å) | 2.752(3) | 2.726(3) |
| Li(1)–N (Å) = *a*/2 | 1.9075(1) | 1.9029(1) |
| Li(2)–N (Å) = *a*/√2 | 2.6975(1) | 2.6912(1) |
| *wR$_p$* | 5.773 | 6.155 |

FeSe (tetragonal phase), FeSe (hexagonal phase), Fe and LiNH/D$_2$ impurities were also modelled. The average refined weight percentages of these phases for the two data sets were FeSe (hexagonal) 10.3%, LiND$_2$ 1.8%, Fe 1.5% and FeSe (tetragonal) 1.2%.

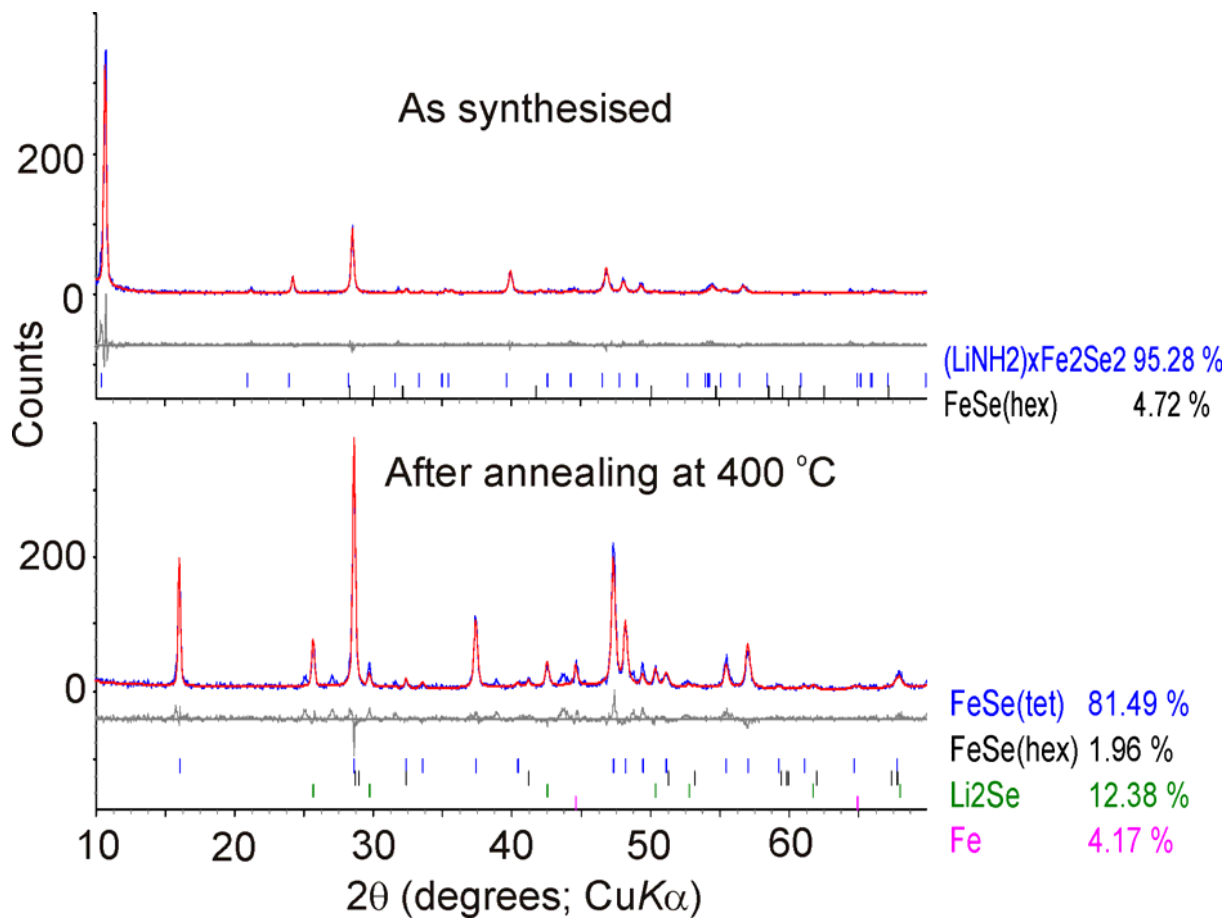

**Figure S2.** Decomposition products of a sample $Li_{0.6(1)}(NH_2)_{0.2(1)}(NH_3)_{0.8(1)}Fe_2Se_2$ after heating at 400 °C under vacuum. The main product is tetragonal FeSe and crystalline $Li_2Se$ and Fe are formed as significant further products in the molar ratio 8 : 2 : 1. The weight fractions of these phases imply a composition "$Li_4Fe_9Se_{10}$" broadly consistent with the composition expected from the synthesis. Some small reflections remain unindexed. The decrease in the amount of hexagonal FeSe is consistent with annealing at 400 °C.

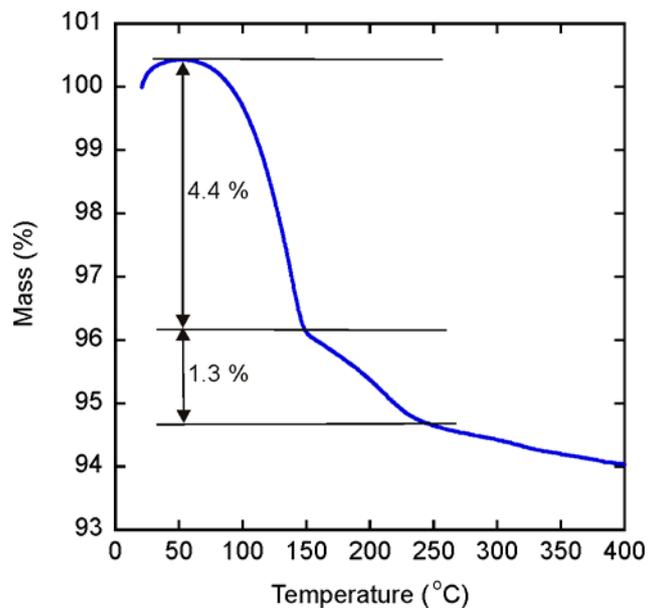

**Figure S3.** Thermogravimetric decomposition of the μSR sample under flowing argon. The mass changes are consistent with the decomposition process suggested by the equation. The small increase in mass at the start of the measurement arises from buoyancy effects in the apparatus.

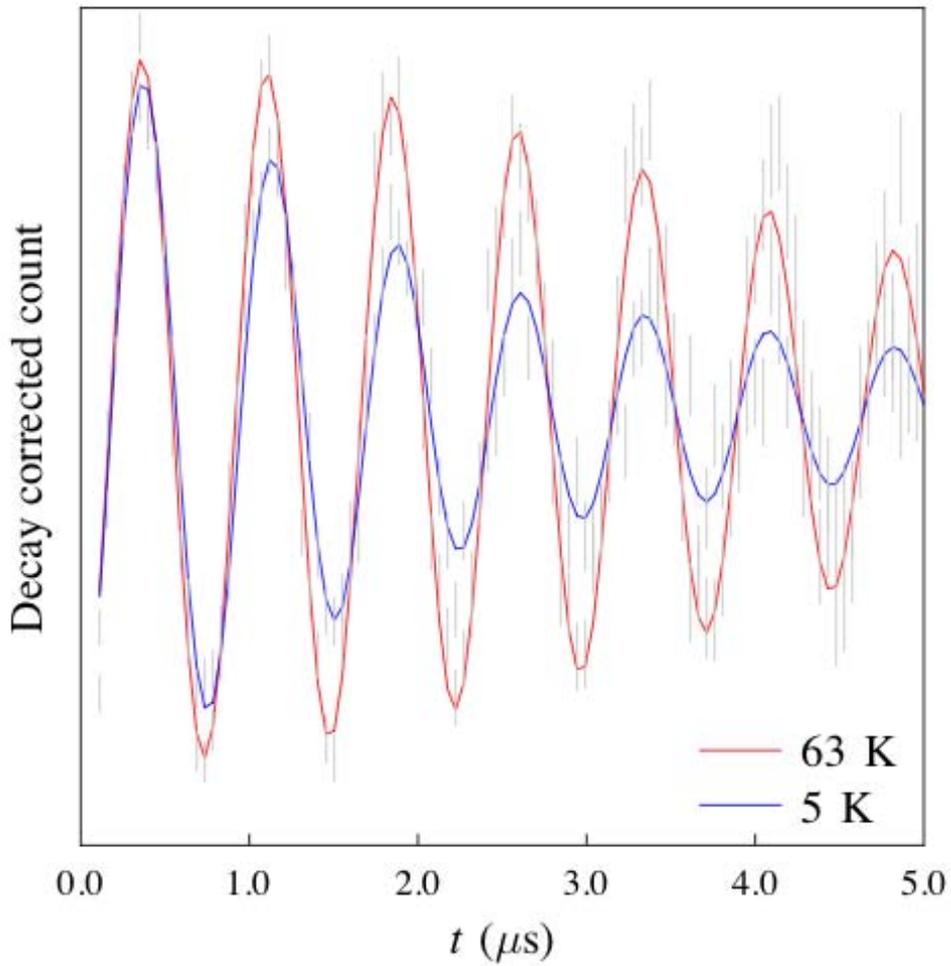

**Figure S4.** Raw muon data for $Li_{0.6(1)}(NH_2)_{0.2(1)}(NH_3)_{0.8(1)}Fe_2Se_2$ in a transverse field of 10 mT at temperatures of 63 K and 5 K, showing the increase in broadening that develops in the superconducting state. The plot shows the counts in a single detector bank, corrected for the muon lifetime decay. The data analysis was carried out using the data from all 64 detectors of the MuSR spectrometer, grouped into 8 detector banks yielding signals with different phases and these were individually fitted. From this analysis the average field and rms width of the field distribution of the magnetic field can be extracted.

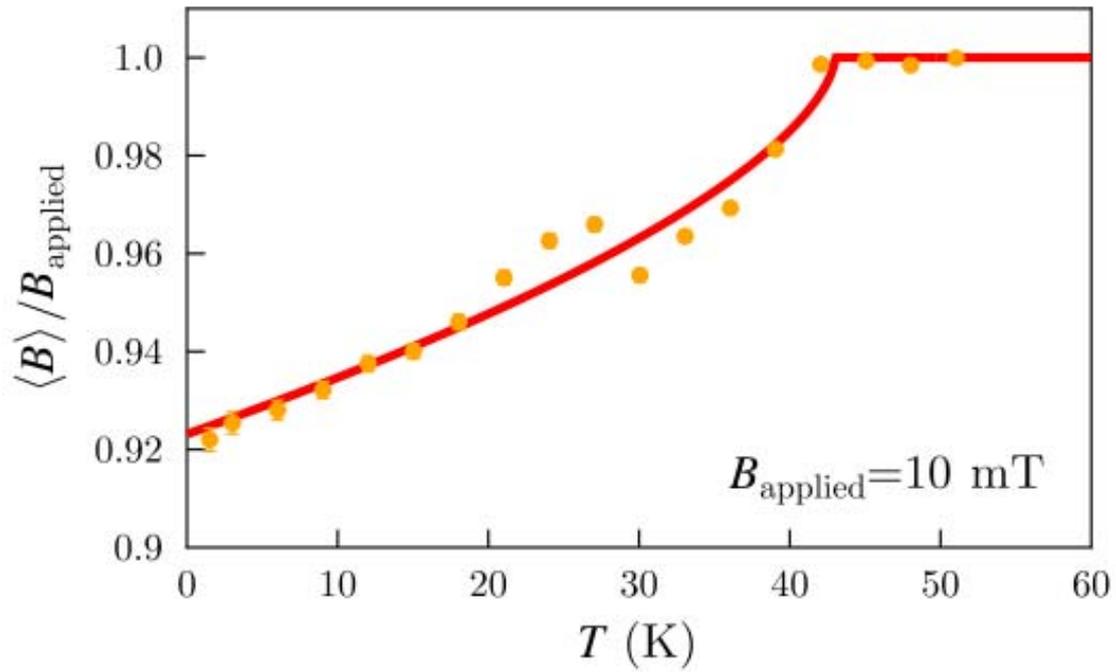

**Figure S5.** The average magnetic field experienced by implanted muons in Li$_{0.6(1)}$(NH$_2$)$_{0.2(1)}$(NH$_3$)$_{0.8(1)}$Fe$_2$Se$_2$ as a function of temperature and normalized by the applied field, showing the diamagnetic shift observed in the superconducting state. These data are extracted from the same fits of the raw data that were used to construct Figure 4(a).

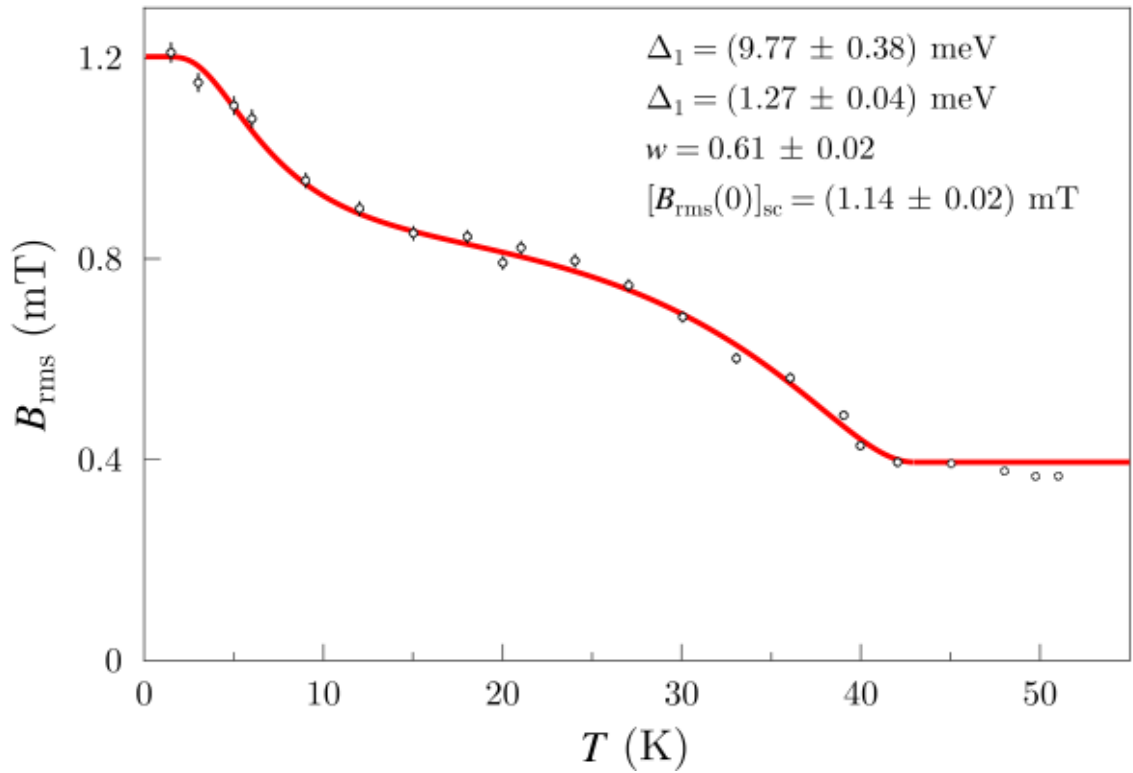

**Figure S6.** The data plotted in Figure 4(a) [main paper] can also be fitted using the expressions used in Ref.25 developed for modelling two-gap superconductors [see also Bussmann-Holder A., Micnas R. and Bishop A. R., Enhancements of the superconducting transition temperature within the two-band model, *Eur. J. Phys. B* 37, 345-348 (2004)]. Such an approach allows an estimate of the size of the gaps (shown in the figure) within the framework of this model. This fitting procedure leads to the same value of the penetration depth given in the main paper.